# Modularity Index Metrics for Java-Based Open Source Software Projects

Andi Wahju Rahardjo Emanuel
Informatics Bachelor Program,
Faculty of Information Technology,
Maranatha Christian University,
Bandung, Indonesia

Retantyo Wardoyo, Jazi Eko Istiyanto,
Khabib Mustofa
Dept. of Computer Science and Electronics,
Universitas Gadjah Mada,
Yogyakarta, Indonesia

*Abstract* — Open Source Software (OSS) Projects are gaining popularity these days, and they become alternatives in building software system. Despite many failures in these projects, there are some success stories with one of the identified success factors is modularity. This paper presents the first quantitative software metrics to measure modularity level of Java-based OSS Projects called Modularity Index. This software metrics is formulated by analyzing modularity traits such as size, complexity, cohesion, and coupling of 59 Java-based OSS Projects from sourceforge.net using SONAR tool. These OSS Projects are selected since they have been downloaded more than 100K times and believed to have the required modularity trait to be successful. The software metrics related to modularity in class, package and system level of these projects are extracted and analyzed. The similarities found are then analyzed to determine the class quality, package quality, and then combined with system architecture measure to formulate the Modularity Index. The case study of measuring Modularity Index during the evolution of JFreeChart project has shown that this software metrics is able to identify strengths and potential problems of the project.

I. INTRODUCTION

Open Source Software (OSS) Projects are gaining popularity these days. They were once only considered as an experimental way of academics and researchers to share the programming experiences, now they become the mainstream software development methodology comparable to those of commercial and proprietary software projects. This movement was initially started by Richard Stallman [33] and Eric Raymond [31]. Some success stories of OSS Projects include Linux Operating System, Apache Web Server, Mozilla Web Browser, LibreOffice, etc. The success of these projects is attributed to many key success factors such as the fact that the developer is the actual user [10], and sound and modular architecture [20][17][11], the existence of communities that support the system development [9], etc. From all these success factors, modularity of the software system is one of the important factors to be examined further in this paper.

Even though there are some proofs of the success of OSS Projects, some facts that many more similar projects are unsuccessful or failed also unavoidable exist [16]. There are some characteristics of OSS Projects that have been identified contributing to such unfruitful result such as no formal means i.e. no project planning [4], poor coding styles of project initiators [13] and poor architectural design [12]. We believe that some new approaches with respect to modularity to counter such problems in OSS Projects are needed. Until now, modularity has been identified as a key success factor of OSS projects, but how to apply modularity, especially from early phase of the project is not yet understood.

This paper presents the formulation of Modularity Index which is the first quantitative software metrics to measure the modularity level in OSS Projects.

This paper is organized as follows: section 2 describes the recent studies in OSS Projects, modularity in OSS Projects and Software Metrics. Section 3 describes the data source of OSS Projects for analysis. Section 4 shows the step by step Modularity Index formulation starts from class level, package level, and system level. The case study of 33 out of 52 versions of JFreeChart projects is shown in section 5. Finally, section 6 describes the conclusion of the paper and future studies of the research.

II. RECENT STUDIES

*A. OSS Projects*

Many web portals have been developed as an incubator for OSS Project's developers to develop and host their projects. These portals are equipped with many development tools and statistics to assist the project initiator or administrator in improving their projects and other interested contributors to join the projects. Some of the popular portals are Sourceforge.net, freshmeat.net, launchpad.net, and Google Code.

The OSS Projects themselves have several distinct characteristics not found in commercial / proprietary software development [10][26], which are:

- The source code of the application is freely available for everybody to download, improve and modify [31].

- People who contribute to the development of the OSS projects are usually forming a group called communities. The recruitment process if this groups are completely voluntary [9]. This communities is an example of true merit-based system of hierarchy [11]

- The development methods of the projects are lacking of formal methodology found in commercially developed software applications [4]. The two most important activities are fixing bugs and adding features [3].

There are already many studies relating to OSS Projects that are classified into three main categories. The first category is the study of large and successful OSS Projects to find their success characteristics such as Debian [32], FreeBSD [12], Apache [27], Open BSD [22], and many more. The second category is the study to find similarities in several OSS Projects such as Apache dan Mozilla [26], 15 OSS Projects [35], and 2 OSS Projects [6]. The last category is the study on the process aspects in OSS Projects such as Requirement Engineering [30], code fault [22], Design Pattern [18], reliability model [37], phase of development [34], and work practice in OSS projects [10].

Current studies about OSS Projects mostly focus on the already successful and large projects that have already established hierarchy and system, while most of the failed and unsuccessful OSS Projects are usually small or medium sized projects [16]. The application of these hierarchy and system in already established projects into small to medium sized projects may not be suitable. In our initial research, we have conducted analysis on more than 130K OSS Projects to find their success factors [15].

*B. Modularity in OSS Projects*

Modularization involves breaking up of an software system into smaller, more independent elements known as module [23]. Booch has defined modularity as the property of a system whose modules are cohesive and loosely-coupled [24]. Fenton stated that modularity is the internal quality attribute of the software system [24]. It is also known that modularity is directly related to software architecture, since modularity is separation of a software system in independent and collaborative modules that can be organized in software architecture [29]. Modular software has several advantages such as maintainability, manageability, and comprehensibility [28]. Moreover, modularity has been identified as one of the key success factors in OSS Projects [20][17][11].

There are five attributes closely related to modularity in software system which are coupling / dependency, complexity, cohesion, and information hiding [21][7]. To have an ideal modular software system, the system should have the following attributes:

- Small size in each module (package) and many modules in the system [36]: each module / package should only responsible for simple feature, and the more complex features should be composed of many of these simple features. The possible software metrics to measure size are NCLOC (non-commenting lines of code), Lines, or Statements.

- Low coupling / dependency [5]: minimization or standardization of coupling / dependency e.g. through standard format i.e. published APIs [2], elimination of semantic dependencies, etc.

- Low complexity: hierarchy of modules that prefers flatter than taller dependency [28][2].

- High cohesion [21]: high integrity of the internal structure of software modules which is usually stated as either high cohesion or low cohesion.

- Open for extension and close to modification [5]: capability of the existing module to be extended to create a more complex module. And avoid changing already debugged code. The creation of new modules should be encourage using available extension and not modifying the already tested module.

Even though modularity is already identified as the key success factor in OSS Projects, the justification for it in large and succesful OSS Projects is purely qualitative. The software metrics attributing to the modularity properties are all separated and not yet integrated into a single measure. This paper will present a single measure called Modularity Index that quantitatively determines the modularity level of OSS Projects.

*C. Software Metrics*

Software metrics are defined as certain values which are expressed in some units attributed to software application [25]. The software metrics are useful in indicate the current state of the software and enable to compare and predict the current achievement of software applications [25]. There are several known software metrics based on its categories [25]:

- Size-related software metrics: NCLOC, Memory footprint, Number of classes / headers, Number of methods, Number of attributes, Size of compiled code, etc.

- Quality-related software metrics: Cyclomatic complexity, Number of states, Number of bugs in LOC, Coupling metrics, Inheritance metrics, etc.

- Process-related software metrics: failed builds, defect per hour, requirement changes, programming time, number of patches after release, etc.

There are currently more than 200 metrics with many different purposes [25], and one of the study by the authors are the statistical analysis of software metrics affecting modularity in OSS Projects [14].

### III. DATA SOURCE OF OSS PROJECTS

The data source of the OSS Projects for the experiment is from the sourceforge.net portal since it is the largest OSS Portal.

*A. Assumptions and Considerations*

There are several consideration and assumption in selecting which OSS Projects to be analyzed, which are:

- The OSS projects are build using Java programming language, and a single package in the project resembles a "module" in modular software system. The addition of package in the software is intended as the addition of new feature in the system.

- The project's size is limited to small-to-medium-sized OSS Projects. The limitation of the size (NCLOC) of OSS Projects being evaluated are 170K. The concept of modularity is a lot easier to comprehend in object-oriented programming language (i.e. C++, Java, etc.) compared to procedural programming (i.e. C, Fortran, etc.), since the concept of module, coupling, cohesion, etc. are more straightforward. Java-based OSS Projects are selected since they are among the mostly popular object oriented programming for developing Open Source Software [16].

- The Projects should already be downloaded more than 100,000 times. This high number of downloads may indicate the "success" of the projects, which in turn may imply modularity traits that already identified as the success factor of OSS Project [20][17][11].

- The source code of the OSS Project is syntax error-free and compile-able. The SONAR tool requires that the source code should be compiled first using compile tool such as maven, or ant. Many of the OSS Projects provides separate binary and source code and it is difficult to create binary directly from the source code due to several reasons such as compile error, build tool configuration error, syntax error, etc.

*B. Selected OSS Projects*

Table 1. shows the list of OSS Projects as a subject for this research. The initial OSS Projects to be evaluated are 209 projects, but only 59 which are suitable to be evaluated using SONAR due to the assumptions and considerations stated in section III.A. There are total 1885 modules / packages being measured from these 59 OSS Projects.

TABLE I. LIST OF 59 SELECTED OSS PROJECTS

| No | Project Name | No | Project Name |
|---|---|---|---|
| 1 | FreeMind | 31 | Jin client for chess servers |
| 2 | jEdit | 32 | SAX: Simple API for XML |
| 3 | TV-Browser - A free EPG | 33 | jKiwi |
| 4 | JFreeChart | 34 | Data Crow |
| 5 | JasperReports - Java Reporting | 35 | Wicket |
| 6 | OpenProj - Project Management | 36 | Cewolf - Chart TagLib Project |
| 7 | HyperSQL Database Engine | 37 | DrawSWF |
| 8 | yura.net | 38 | c3p0:JDBC DataSources / Resource Pools |
| 9 | JabRef | 39 | JavaGroups |
| 10 | FreeCol | 40 | OmegaT - multiplatform CAT tool |
| 11 | jTDS - SQL Server and Sybase JDBC driver | 41 | FreeGuide TV Guide |
| 12 | Torrent Episode Downloader | 42 | Eteria IRC Client |
| 13 | FindBugs | 43 | MeD's Movie Manager |
| 14 | PMD | 44 | subsonic |
| 15 | JGraph Diagram Component | 45 | kXML |
| 16 | ANts P2P | 46 | Jaxe |
| 17 | Paros | 47 | The JUMP Pilot Project |
| 18 | ProGuard Java Optimizer and Obfuscator | 48 | Aglet Software Development Kit |
| 19 | TripleA | 49 | Antenna |
| 20 | JSch | 50 | CBViewer |
| 21 | Jajuk | 51 | Sunflow Rendering System |
| 22 | FreeTTS | 52 | Thingamablog |
| 23 | A Java library for reading/writing Excel | 53 | BORG Calendar |
| 24 | checkstyle | 54 | Directory Synchronize Pro (DirSync Pro) |
| 25 | httpunit | 55 | Java Treeview |
| 26 | JMSN | 56 | Java Network Browser |
| 27 | PDFBox | 57 | Red Piranha |
| 28 | JBidwatcher | 58 | Cobertura |
| 29 | JTidy | 59 | Jake2 |
| 30 | Jena | - | - |

*C. Steps*

In order to be able to analyze these OSS Projects, there are some steps being performed, which are:

- Compile the source code using available build tool (Ant or Maven2).

- Execute maven2 script to start analyze the OSS Projects using SONAR tool.

- Creating custom portal to perform the required analysis.

- Analyze and find the correlation and similarities of all the projects such as using scatter graph, least square fit, histogram, etc.

IV. MODULARITY INDEX FORMULATION

The formulation of modularity index will start from the class level, then move up to the package level, and finally concluded in the system level.

*A. Class Level Modularity*

There are four software metrics that determine the level of modularity in class level, which are:

- Size Metrics which consists of: NCLOC, Lines, and Statements. NCLOC is the number of non-commenting lines of code. The selection of NCLOC will also represent the other size metrics [14].

- Cohesion: LCOM4 or Lack of Cohesion Method version 4, this version is better for object oriented programming such as Java as proposed by Hitz and Montazeri [19] which is the improvement of LCOM1 Chidamber and Kemerer [8].

- Complexity: McCabe's Cyclomatic Complexity [22] is one example of complexity metrics that widely used.

Our previous paper have shown that the size metrics and complexity metrics are highly related so this metrics may be ignored [14].

- Functions: the number of functions / methods in the class. This may indicates the complexity

*1) NCLOC:* Figure 1 shows the histogram of the class vs. NCLOC of the all OSS Projects being evaluated. The value of NCLOC peaked at 50 with the histogram before the peak resembles linear straight line and after the peak resembles inverse polynomial line. The value of approximation of both lines are shown in the Fig.1.

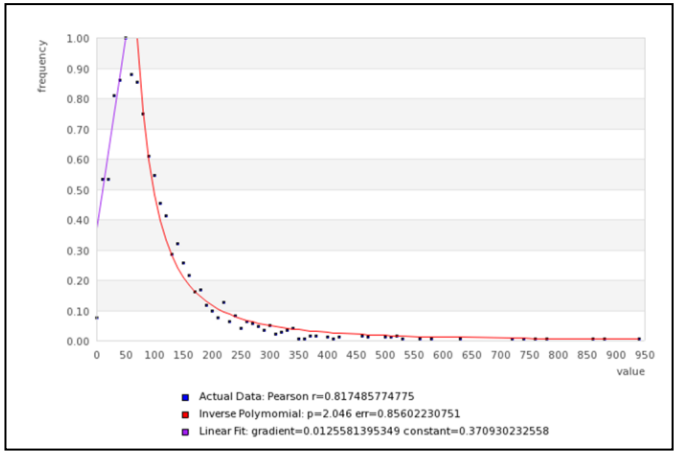

Figure 1. Histogram of Classes vs. NCLOC

If $LOC_Q$ is defined as the normalized value of the quality of NCLOC, so the formula of $LOC_Q$ are:

$$LOC_Q = 0.0125 \times NCLOC + 0.375 \text{ for } NCLOC \leq 50 \quad (1)$$

$$LOC_Q = (NCLOC - 50)^{-2.046} \text{ for } NCLOC > 50 \quad (2)$$

Where:

$LOC_Q$ = NCLOC Quality Value

NCLOC = NCLOC Value

Note: the value of constant in formula (1) is adjusted from 0.371 into 0.375 to achieve the maximum value of 1 at NCLOC = 50.

*2) Number of Functions:* Figure 2 shows the histogram of classes vs. functions of all OSS Projects being evaluated. The peak value is 4.83 (rounded up into 5). Similar to class vs. NCLOC, the values before the peak resembles a straight line and after the peak resembles an inverse polynomial line with the approximation of both lines shown in the Fig.2.

$F_Q$ is defined as the normalized value of function's quality, it can be formulated as follows:

$$F_Q = 0.172 \times F + 0.171 \text{ for } F \leq 5 \quad (3)$$

$$F_Q = (F - 4.83)^{-2.739} \text{ for } F > 5 \quad (4)$$

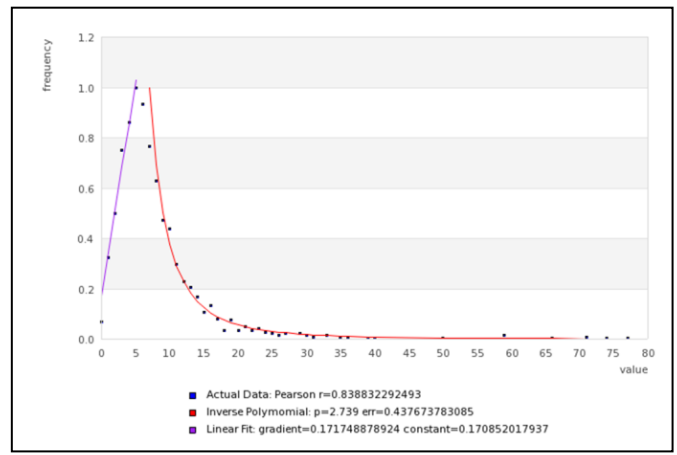

Figure 2. Histogram of Classes vs. Functions

Where:

$F_Q$ = Function Quality Value

F = Number of Function

*3) Cohesion:* Cohesion is determined by the value of LCOM4. The ideal value is 1 which means that the class is highly cohesive. Higher value of LCOM4 indicates the degree of needed separation of classes into smaller classes.

$$LCOM4 \geq 1 \quad (5)$$

Where:

LCOM4 = Class Cohesion Value

*4) Class Quality Formulation:* Integrating all above measures into a single normalized value, the formulation of class quality or $C_Q$ are:

$$C_Q = \frac{LOC_Q + F_Q}{2 \times LCOM4} \quad (6)$$

Where:

$C_Q$ = Class Quality Value

$LOC_Q$ = NCLOC Quality Value

$F_Q$ = Function Quality Value

LCOM4 = Class Cohesion Value

*B. Package Level Modularity*

Package Quality or $P_Q$ is the quality of individual package. Since in a single package there are many classes and there is no similarities found the the optimal number of classes in each package, so the Package Quality is determined by the average Class Quality or stated as:

$$P_Q = avg(C_Q) \quad (7)$$

Where:

$P_Q$ = Package Quality Value

$C_Q$ = Class Quality Value

### C. System Level Modularity

$S_A$ is a normalized value (with maximum value of 1) which determine the value of software architecture. The factors that influence this value are Package Cohesion (relationship among classes within package) and Package Coupling (relationship among classes from different packages). The principle used here is "*Maximize Cohesion and Minimize Coupling*" which becomes a widely known principle in building a good software system. The form of formulation is based on presentation titled "*Software Architecture Metrics*" by Ammar et. al [1], with the difference is that instead of using entropy approaches, this formulation is using the actual value of dependencies in determining the value of Package Cohesion and Package Coupling.

$$S_A = \sqrt{\frac{\sum_{i=1}^{d} C_{ii}^2}{\sum_{i=1}^{d}\sum_{j=1}^{d} C_{ij}^2}} \qquad (8)$$

Where:

 $C_{ii}$ = Package Cohesion

 $C_{ij}$ = Package Cohesion + Package Coupling

 (if i=j is Package Cohesion,

 if i ≠ j is Package Coupling)

 d = number of package

### D. Formulation of Modularity Index

Finally, the formulation of Modularity Index is the product of SA and the sum of all package quality in the software system as stated in the following formula:

$$M_I = S_A \times \sum_{i=1}^{j} P_{Qi} \qquad (9)$$

Where:

 $M_I$ = Modularity Index

 $S_A$ = Software Architecture Value

 $P_{Qi}$ = Package Quality of Package i

The proposed modularity index is a quality metrics will have the following properties:

- It has no upper bound: the value of modularity index increases as the number of module / package increases.

- The value of modularity index, especially the value of $S_A$ depends on how the packages are coupled to each other. The limitation of connection of packages to only itself (package cohesion) or to only some dedicated packaged (e.g APIs, proxy, etc.) will improve the value of $S_A$.

## V. CASE STUDY: JFREECHART

JFreeChart is a free 100% Java chart library that makes it easy for developers to display professional quality charts in their applications (http://www.jfree.org/jfreechart) . This projects is one of the 59 OSS Projects used for modularity index formulation. For this case study, this project is chosen because:

- High $S_A$ value (more than 0.7 since version 0.9.21)

- Relatively large number of packages (more than 30)

There are 52 versions available from the project's site, but only 33 are able to be analyzed using SONAR tool and being measured. The results are show in the following Fig.3.

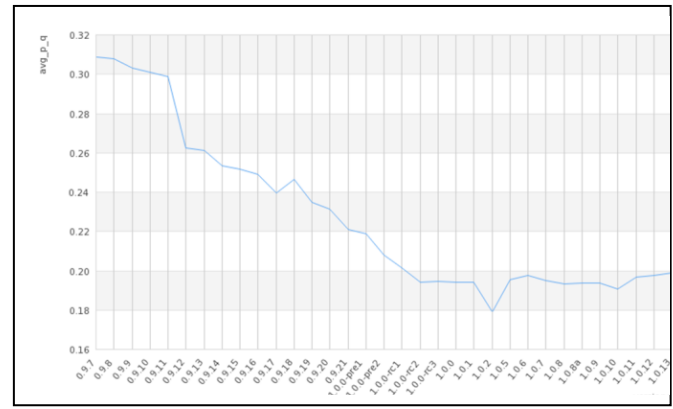

Figure 3. Average $P_Q$ in 33 versions of JFreeChart

Fig. 3 above shows that the average package quality of the JFreeChart over 33 versions are decreasing consistently. This indicates the problem in the quality of each classes in each packages, such as:

- increasing size of NCLOC in each class.

- increasing number of functions in class.

- decreasing number of LCOM4 (Cohesion Metrics) in class.

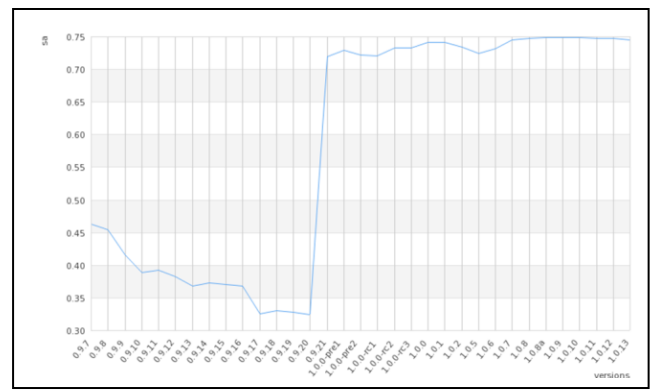

Figure 4. SA value in 33 versions of JFreeChart

Fig. 4 above shows that the structure of software architecture is improving. After consistent decrease in SA value in early versions of the system, there seems to be significant effort conducted before the release of version 1.0.0 started from version 0.9.21. The system from version 0.9.21 onward showing high number of SA.

The modularity index itself is shown in Fig. 5. The figure is showing improvement by the factor of two from early versions (until version 0.9.20) and late versions (version 1.0.5 onwards). There are significant jump in the value of modularity index from version 0.9.21 until version 1.0.2 indicating the period of major restructuring of the system before the release of milestone version 1.0.0.

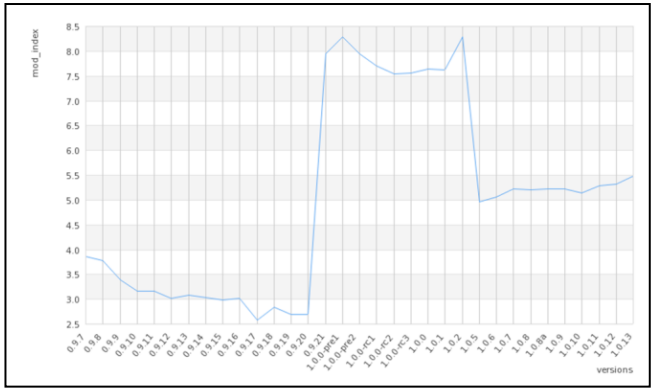

Figure 5. Modularity Index in 33 versions of JFreeChart

It can be seen from above case study that Modularity Index and its components ($P_Q$ and $S_A$) are able to point the strength and potential problems in the development of JFreeChart OSS Projects. This information may give a valuable insight to the initiator and developers of the project in improving their project.

## VI. CONCLUSION

Open Source Software (OSS) Projects are now gaining popularity and becoming one alternatives in developing software. Despite the the many success story of OSS Projects such as Apache, Mozilla, etc., the fact the many more of these projects that are failed needs are alarming. Some studies have identified that modularity is one of the key success factors of OSS Projects and authors believe that implementing modularity approach since early start of the project will increase the success of the project. This paper presents the first quantitative measure of modularity for Java-based OSS Projects called modularity index.

The formulation of modularity index are performed by analyzing the software metrics attributing to modularity of 59 Java-based OSS Projects from sourceforge.net which have been downloaded more than 100K times. By analyzing the similarity of these projects from class level, package level, and system level, the modularity index are formulated. As the validation of the software metrics, 33 out of 52 versions of JFreeChart OSS projects are analyzed using this metrics and the metrics are able to identify the strength and potential problems of the project.

Future study relating to this metrics involve further validation and integration into a framework called modularity framework in which the measurement of Modularity Frameworks will generate recommendations for improvement during OSS project's development. The integration of the software metrics into a web-based IDE will provide useful tool for project initiators and developers in improving their OSS Projects.

ACKNOWLEDGMENT

The authors would like to thank Maranatha Christian University (http://www.maranatha.edu) that provides funding for the research, and the Department of Computer Science and Electronics, Universitas Gadjah Mada (http://mkom.ugm.ac.id) that provides technical support for the research.

AUTHORS PROFILE

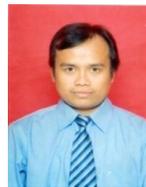

Andi Wahju Rahardjo Emanuel is a Full Time Lecturer at the Bachelor Informatics Program, Faculty of Information Technology, Maranatha Christian University in Bandung, Indonesia. He is graduated as BSEE in Purdue University, Indiana, USA in 1996 and MSSE in The University of Melbourne in 2001. He is currently taking his Doctoral Program at the Department of Computer Science and Electronics, Gadjah Mada University in Yogyakarta, Indonesia

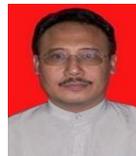

Retantyo Wardoyo is an Associate Professor at the Department of Computer Science and Electronics, Universitas Gadjah Mada in Yogyakarta, Indonesia. He is graduated as Bachelor of Mathematics in Gadjah Mada University, Indonesia . He received his M.Sc in Computer Science in University of Manchester, UK and received his PhD in Computation in University of Manchester Institute of Science and Technology, UK.

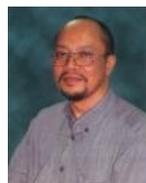

Jazi Eko Istiyanto is a Professor and Head of the Department of Computer Science and Electronics, Universitas Gadjah Mada in Yogyakarta, Indonesia. He is graduated as Bachelor of Physics in Gadjah Mada University, Indonesia. He gets his Postgraduate Diploma (Computer Programming and Microprocessor), M.Sc (Computer Science) and PhD (Electronic System Engineering) at University of Essex, UK.

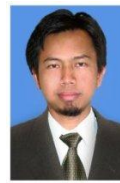

Khabib Mustofa is an Assistant Professor at the Department of Computer Science and Electronics, Universitas Gadjah Mada in Yogyakarta, Indonesia. He is graduated as Bachelor of Computer and Master of Computer at Gadjah Mada University, Indonesia. He receives his Dr. Tech in Computer Science at The Vienna University of Technology, Austria.